\documentstyle[12pt,epsf,epsfig,wrapfig]{article}

\RequirePackage{amssymb}

\newcommand{\rar}{\rightarrow}
\newcommand{\pdup}{p_\uparrow}

\newcommand{\ppdup}{p + \pdup \rar \pi^0 + X}
 
\newcommand{\pimp}{\pi^- + \pdup \rar \pi^0 + X}

\newcommand{\xf}{x_{\mathrm F}}
\newcommand{\pdupp}{\pdup + p \rar \pi^0 + X}
\newcommand{\PRD}[1]{{Phys.\ Rev. D}\ {\bf #1}}
\newcommand{\PRL}[1]{{Phys.\ Rev.\ Lett.}\ {\bf #1}}
\newcommand{\PLB}[1]{{Phys.\ Lett. B}\ {\bf #1}}

\begin{document}
\begin{center}
{\bfseries RECENT RESULTS FROM PROTVINO POLARIZED EXPERIMENT PROZA-M}
\vskip 5mm

\underline {V.V. Mochalov$^{1 \dag}$}, N.S.~Borisov$^2$, 
A.M.~Davidenko$^1$,  A.N.~Fedorov$^2$, V.N.~Grishin$^1$,  V.Yu.~Khodyrev$^1$, 
V.I.~Kravtsov$^1$, A.A.~Lukhanin$^3$, V.N.~Matafonov$^2$, 
Yu.A.~Matulenko$^1$, V.A.~Medvedev$^1$, Yu.M.~Melnick$^1$, 
A.P.~Meschanin$^1$, D.A.~Morozov$^1$, A.B.~Neganov$^2$, L.V.~Nogach$^1$, 
S.B.~Nurushev$^1$, Yu.A.~Plis$^2$, A.F.~Prudkoglyad$^1$, P.A.~Semenov$^1$,
K.E.~Shestermanov$^1$, \fbox{V.L.~Solovianov$^1$},L.F.~Soloviev$^1$, 
Yu.A.~Usov$^2$, A.N.~Vasiliev$^1$, A.E.~Yakutin$^1$ \\

\vskip 5mm
{\small $^1$ {\it Institute for High Energy Physics, Protvino, Russia}\\
{\it Joint Institute for Nuclear Research, Dubna, Russia}\\
{\it Kharkov Physical Technical Institute, Kharkov, Ukraine}\\
$\dag$ {\it
E-mail: mochalov@mx.ihep.su}}
\end{center}

\begin{abstract}
Single Spin Asymmetry  $A_N$ (SSA) was measured at the Protvino 70~GeV
accelerator. Asymmetry in the reaction $\ppdup$ is close to zero 
within error bars at small $|\xf|$ and $p_T<1.5$~GeV/c. 
SSA in the reaction $\pimp$ at polarized target fragmentation region
equals to $(-13.8 \pm 3.8)\%$ at $|x_f|>0.4$. 
There is an indication that the asymmetry
begins to rise up at the same  centre of mass system pion energy.
\end{abstract}

\vskip 5mm

Large polarization effects were found during last few decades.
Single Spin Asymmetries were observed to be of order 
20--40\%  while Perturbative Quantum Chromodynamics (pQCD) makes 
a qualitative prediction that the single-spin transverse effects 
should be very small due to the helicity conservation \cite{kane}.  
Here we present new SSA measurements carried out at the 
70~GeV Protvino accelerator in the reaction $\ppdup$ at 70~GeV in 
the central region ($\xf \approx 0$) and in the reaction $\pimp$ in 
the polarized target fragmentation region at 40~GeV. 

\subsection*{Asymmetry in the reaction $\ppdup$ at 70~GeV.}

The goal of our measurements was to measure asymmetry 
in the reaction $\pimp$  and to compare it with the previous
contradictory results. 

\begin{wrapfigure}{R}{11cm}
\mbox{\epsfig{figure=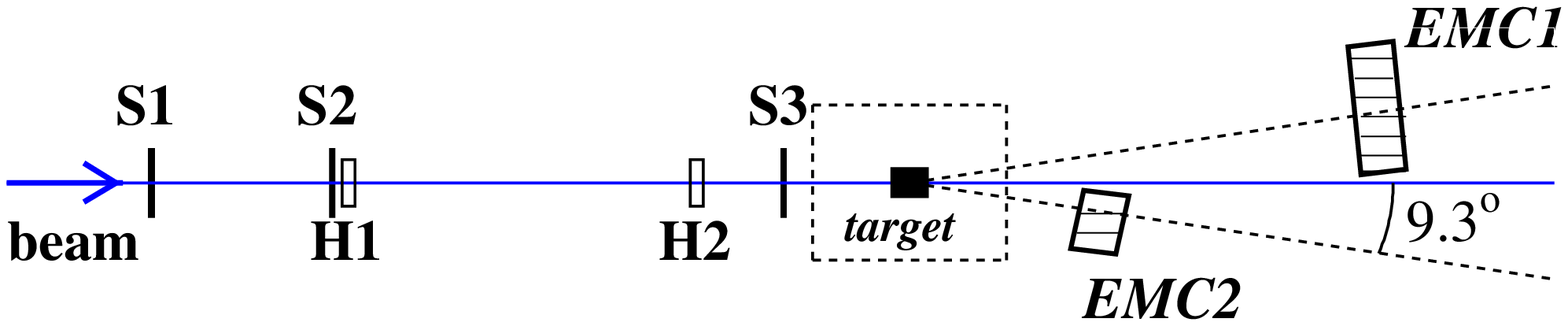,width=10.8cm,height=3.4cm}}\\
{\small {{\bf Figure 1.} Experimental Setup PROZA-M. 
S1-S3 -- trigger scintillation counters; H1-H2 -- hodoscopes; 
$EMC1$ and $EMC2$ -- electromagnetic calorimeters;
$target$ -- polarized target.}}
\end{wrapfigure}

Asymmetry in the reaction $\ppdup$ in the central 
region was measured earlier in two experiments. 
Experiment at 24~GeV (CERN) manifested large SSA in 
inclusive $\pi^0$ production \cite{dick24}, 
while E704 (FNAL, 200~GeV) claimed zero asymmetry \cite{E704center}. 

\begin{wrapfigure}{R}{7.9cm}
\mbox{\epsfig{figure=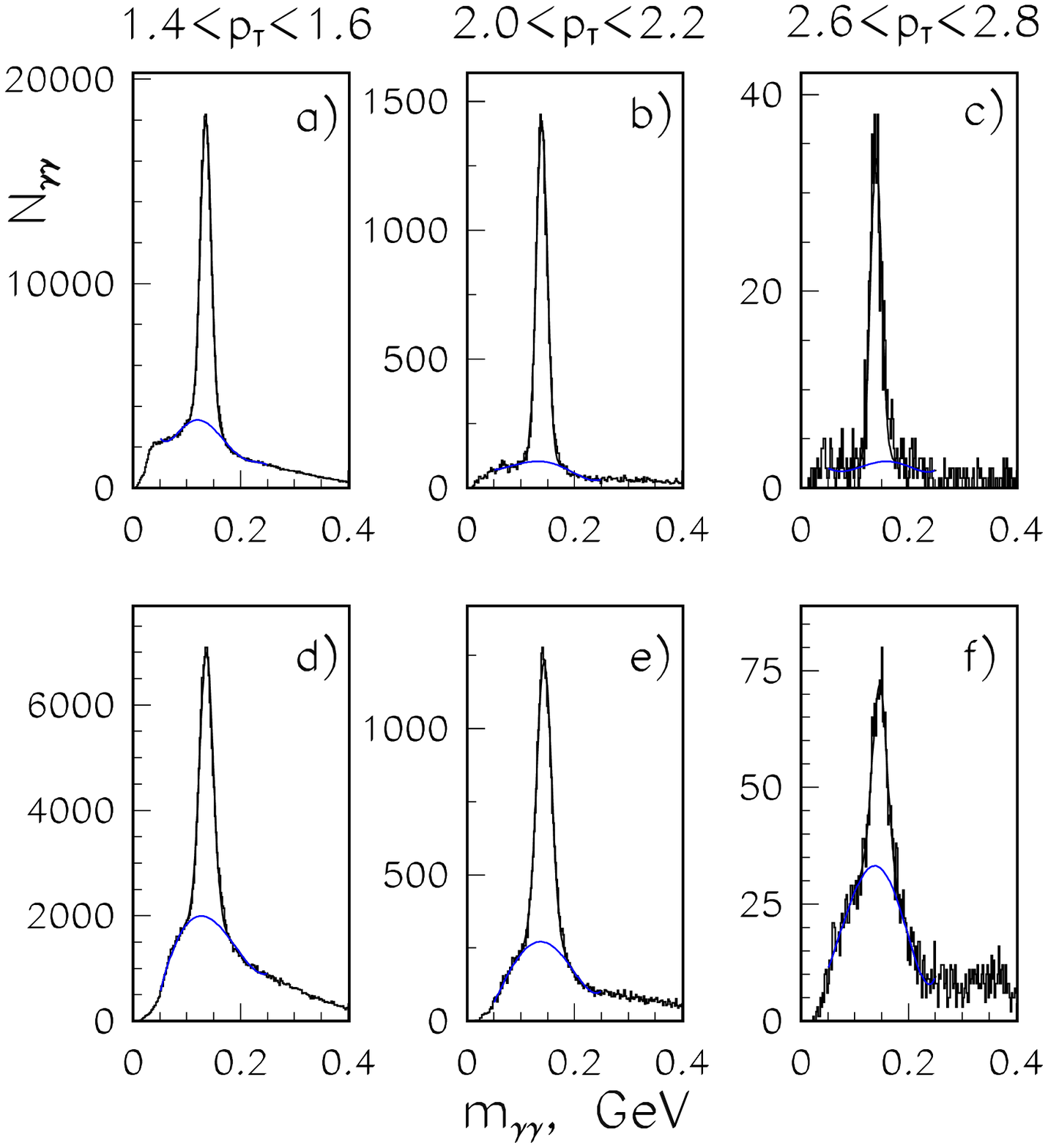,width=7.8cm,height=7.8cm}}\\
{\small {\bf Figure 2.} Mass spectra for EMC1 ($a-c$) and EMC2 ($d-f$) for
different $p_T$ intervals.}
\vspace*{0.3cm}
\end{wrapfigure}

PROZA-M experiment  observed significant effects 
in the reaction $\pimp$  at 40~GeV (Protvino, \cite{protv}). 
New measurements at the Protvino U-70 accelerator facility were carried
out. 70~GeV protons were extracted from the accelerator vacuum chamber
using a bent crystal. The experimental setup is presented in {\bf Fig.~1}.

Three scintillation counters S1-S3 were
used for zero level trigger with a coincidence from two 
hodoscopes  H1-H2 (two subplanes each).  
$\gamma$-quanta were detected by two 
electromagnetic calorimeters $EMC1$ and $EMC2$ (arrays of 480 and 144
lead-glass cells) placed at 7 and 2.8~m downstream the frozen
polarized target with  average polarization of 80\%. 
First level trigger on transverse energy was set up to be
independent for each detector.

\begin{wrapfigure}{L}{6.25cm}
\hspace*{0.1cm} 
\mbox{\epsfig{figure=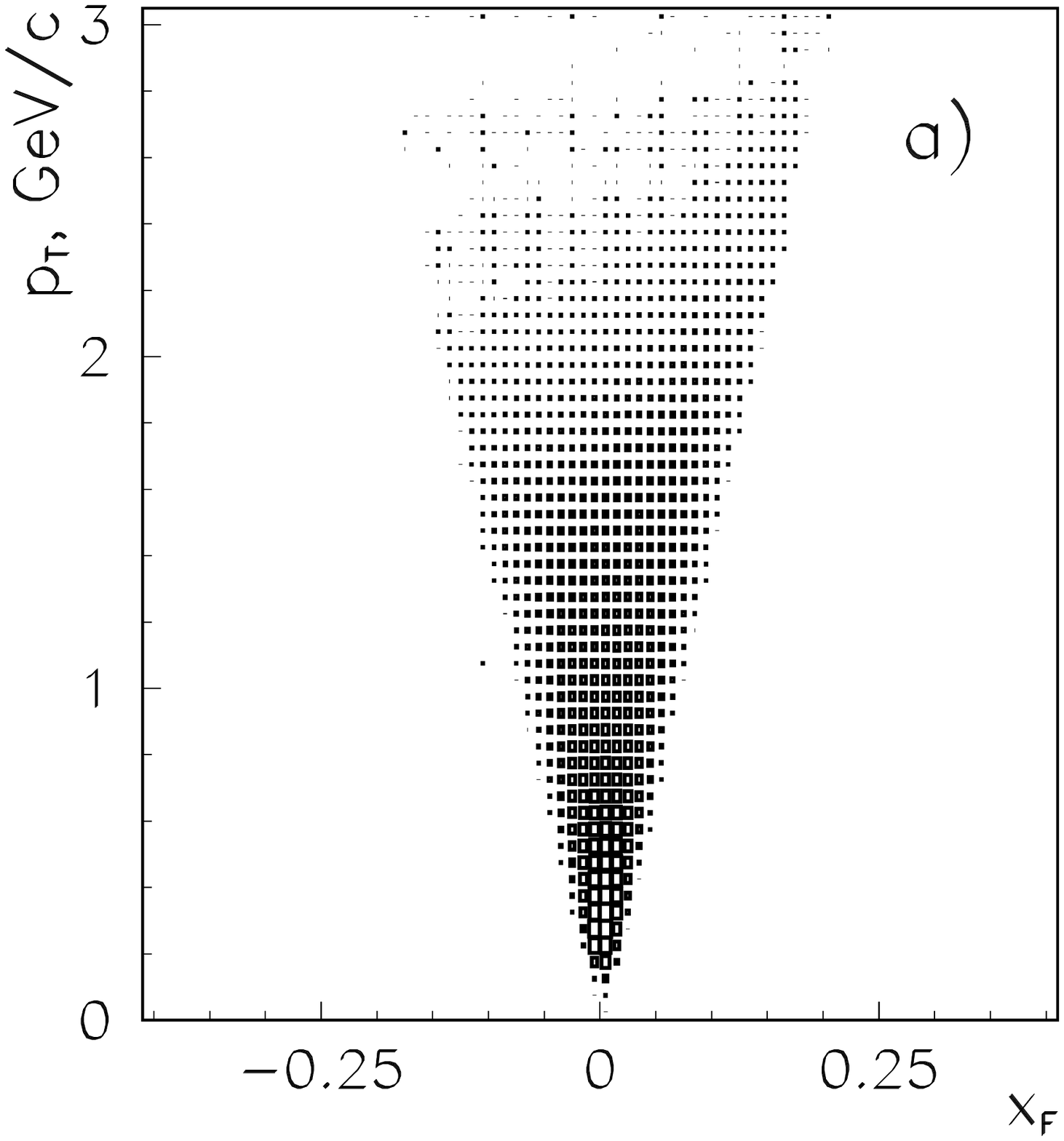,width=5.4cm,height=5.4cm}}\\
\hspace*{0.1cm} 
\mbox{\epsfig{figure=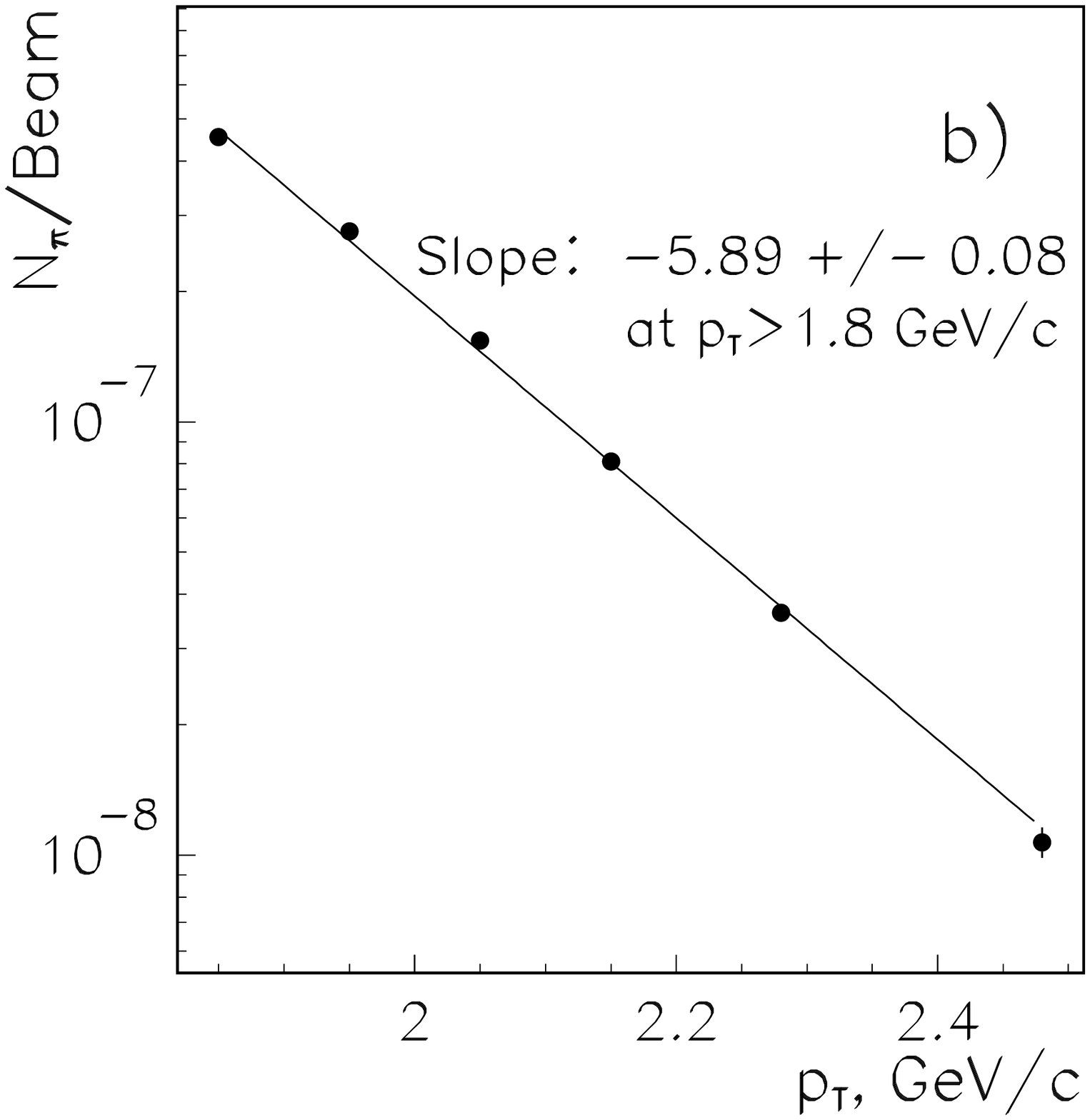,width=5.4cm,height=5.4cm}}\\
{\small {\bf Figure 3.} (a) Two-dimensional $\pi^0$ 
distribution on $p_t$ and $\xf$ and
(b) Relative $\pi^0$ cross-section ($N_{pions}/Beam$).}
\end{wrapfigure}

We were able to detect $\pi^0$ till $p_T=3.0$~GeV/c using  
specially developed algorithm for overlapping showers 
reconstruction \cite{lednev}. 
$\pi^0$ mass resolution was 10~MeV/$c^2$ for EMC1 and from 12 to
16~MeV/$c^2$ for EMC2 ({\bf Fig.~2}).

Angle $9.3^{\circ}$ in the laboratory frame corresponds to 
$90^{\circ}$ in the center of mass system (c.m.s.). Kinematical
characteristics of detected $\pi^0$'s ( $p_T$ vs $\xf$ bi-plot) 
are presented in ({\bf Fig.~3a}). $|\xf|$ did not 
exceed 0.15  and the distribution  was symmetrical on $\xf$. 

The slope of relative $\pi^0$-cross-section 
presented on {\bf Fig.~3b} was used as an 
additional cross-check of data quality.
The result is in a good agreement 
with the previous  charged pion invariant cross-section measurements. 
Exponential constant  $\alpha=-5.89 \pm 0.08$, 
while FODS experiment (Protvino) announced $\alpha=-5.68 \pm 0.02$ 
for $\pi+$'s and $\alpha=-5.88 \pm 0.02$ for $\pi^-$-mesons \cite{FODScross}.  

The asymmetry $A_N$  was defined as

\begin{flushright}
\begin{displaymath}
A_N=\frac{D}{P_{target}}\cdot A_N^{raw} =
\frac{D}{P_{target}}\cdot 
\frac{n_{\downarrow}-n_{\uparrow}}{n_{\downarrow}+n_{\uparrow}}
\end{displaymath}
\end{flushright}

\noindent for EMC2 and had the opposite sign for EMC1. $D$ is the polarized 
target dilution factor, $P_{target}$ -- average target polarization,
$n_{\downarrow}$ and  $n_{\uparrow}$ -- normalized numbers of $\pi^{0}$-mesons
for opposite target polarizations.

\begin{wrapfigure}{R}{8.4cm}
\centering
\vspace*{-0.6cm}
\mbox{\epsfig{figure=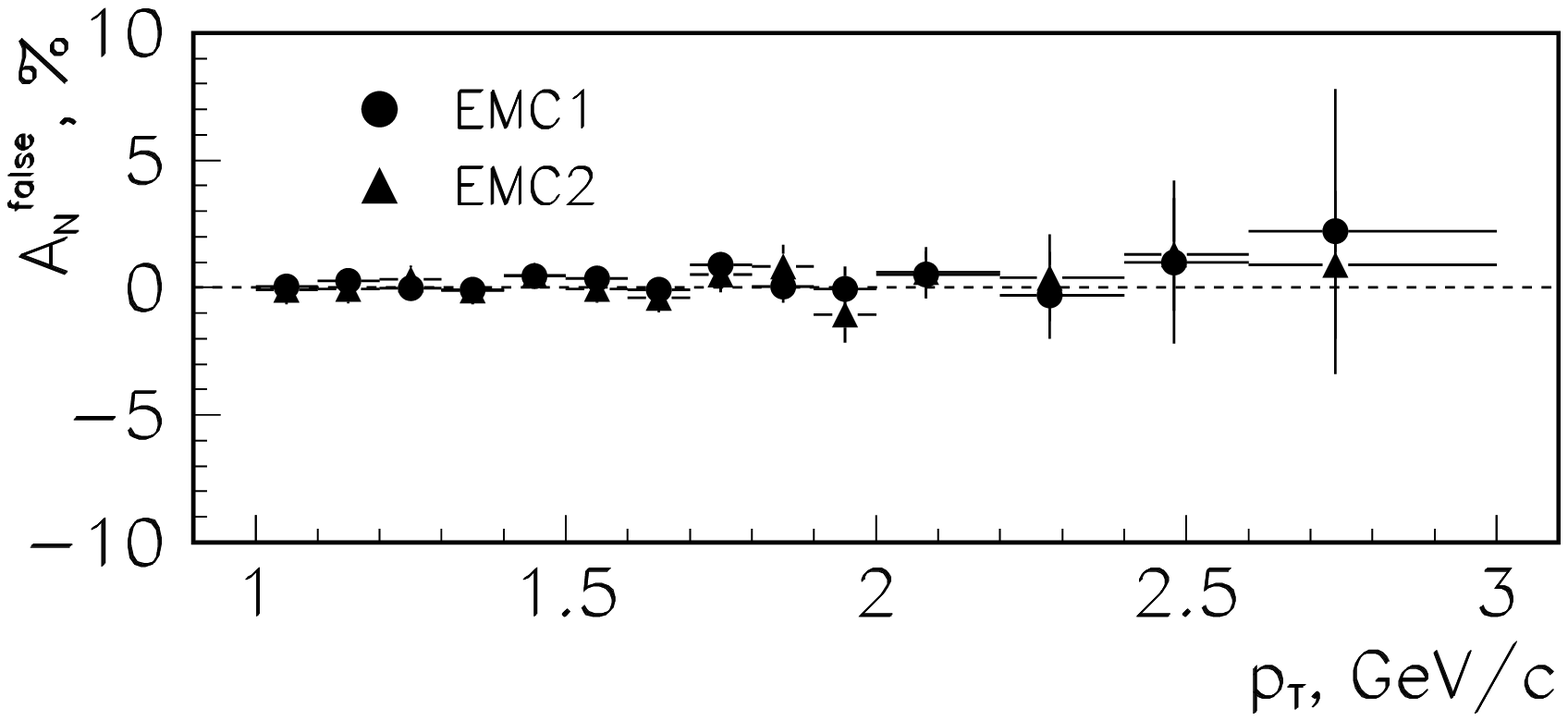,width=8.2cm,height=4.8cm}}\\
\vspace*{-1.cm}
\mbox{\epsfig{figure=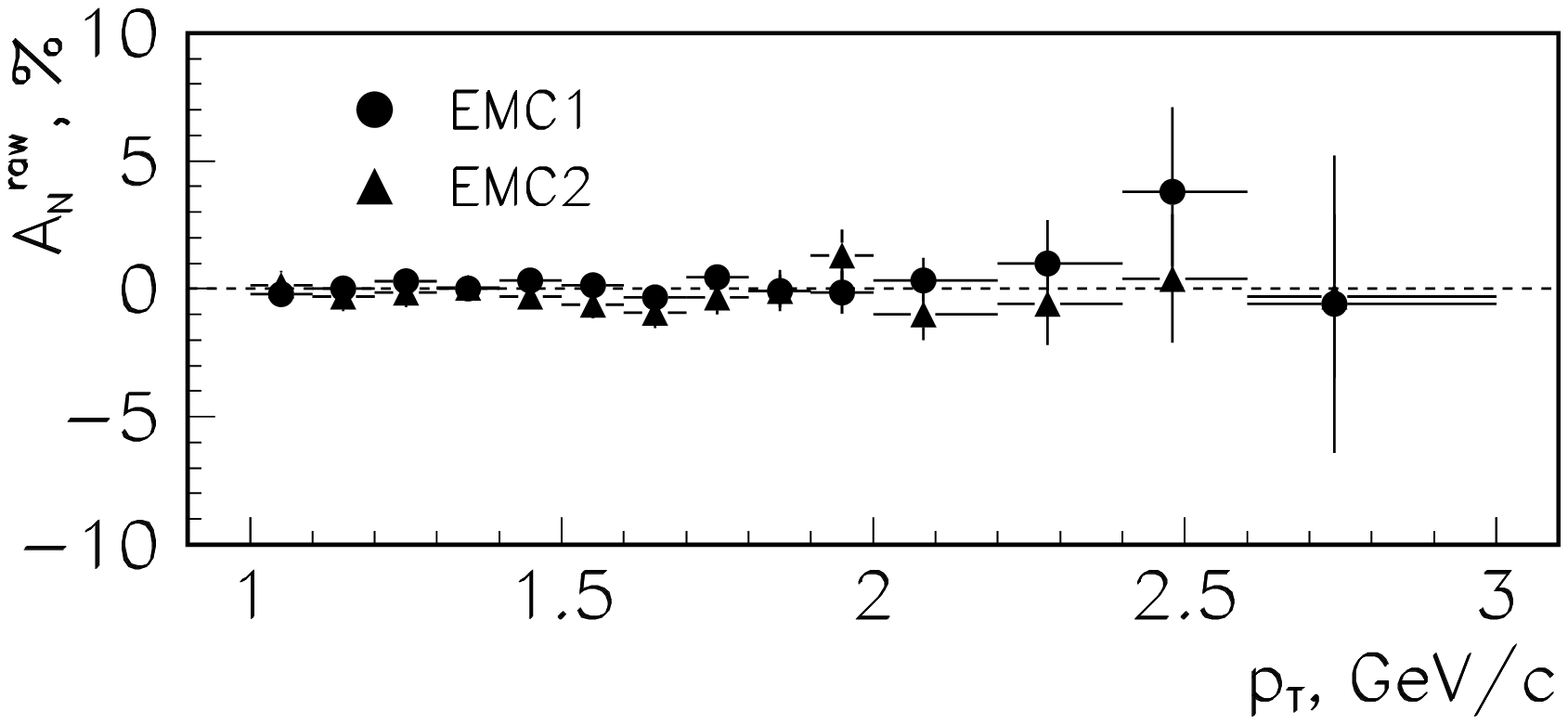,width=8.2cm,height=4.8cm}}\\
{\small {\bf Figure 4.} Raw false asymmetry (top). Asymmetry for
two detectors (bottom)}
\end{wrapfigure}

False asymmetry $A_N^{false}$ appears mainly due to the instability of the 
calorimeters energy scale. We recalibrated both calorimeters independently 
using $\pi^0$ mass with an accuracy of 0.15\%, which  corresponded
to raw false asymmetry value of 0.3\% and $A_N^{false} \approx 3$\% 
taking into account the dilution factor divided by target polarization
($D/P_{target} \approx 10$).  

The false asymmetry was investigated 
by dividing statistics with the
same target polarization into two data samples 
and calculating the asymmetry for them. $A_N^{false}$
was zero within error bars ({\bf Fig4}, top). 
We also compared the asymmetry for the two detectors and 
did not find any difference ({\bf Fig4}, bottom).

The final result for both detectors is presented in 
{\bf Fig5a}. The asymmetry is zero within error bars. 
Our result is in a good agreement with the FNAL data at 200~GeV 
\cite{E704center} 
and contradicts to the previous CERN measurements at 24~GeV\cite{dick24}.

\begin{wrapfigure}{L}{8cm}
\centering
\vspace*{-0.2cm}
\mbox{\epsfig{figure=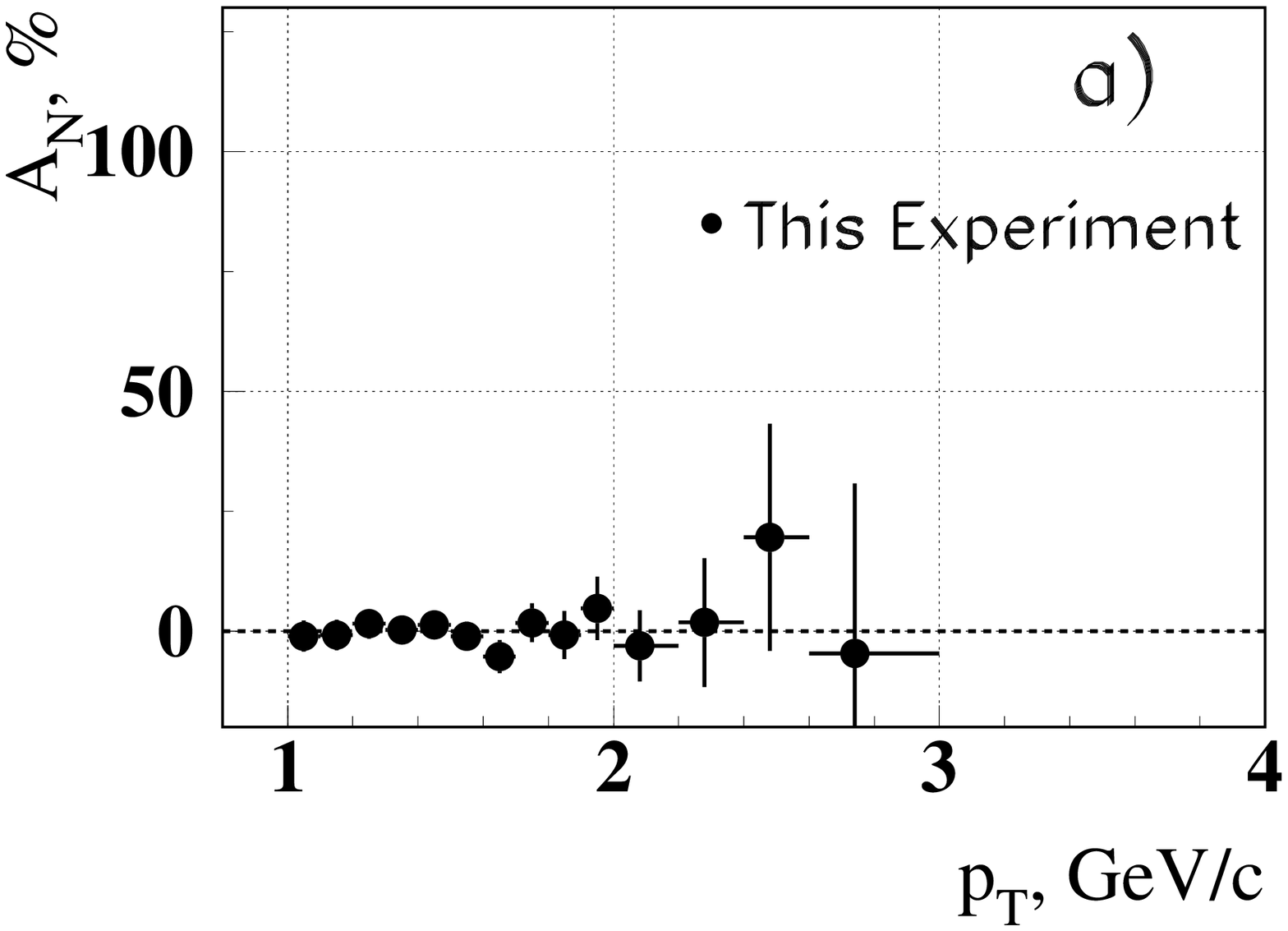,width=7.8cm,height=4.4cm}}\\
\vspace*{-0.4cm}
\mbox{\epsfig{figure=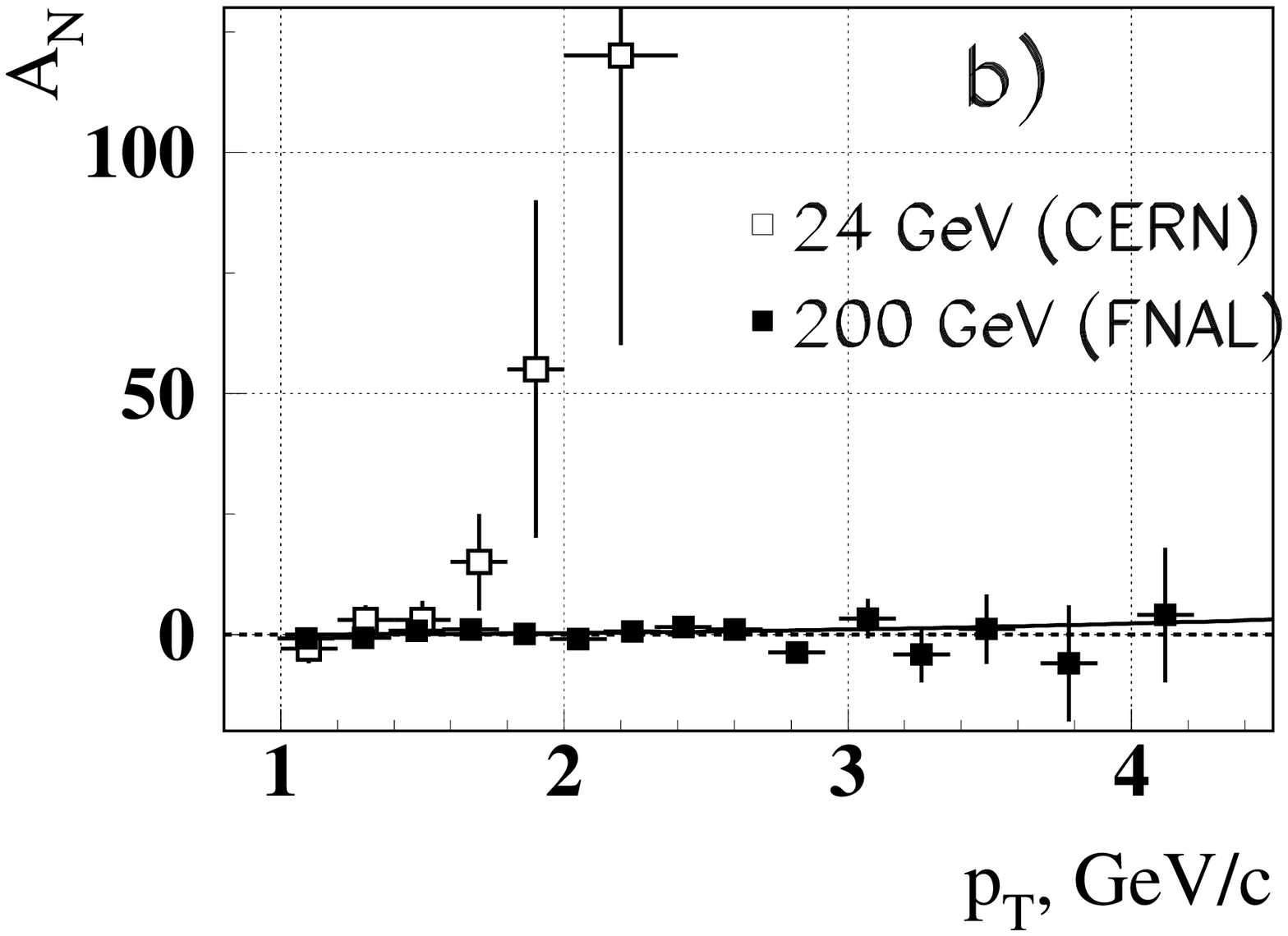,width=7.8cm,height=4.4cm}}\\
{\small {\bf Figure 5.} (a) Summed $A_N$ for two detectors. 
(b) Asymmetry at 24~GeV \cite{dick24} and 200~GeV\cite{E704center}
at central region; Anselmino  calculations \cite{anselm98} 
for 200~GeV and $x_{\mathrm F}=0$ are drawn.}
\end{wrapfigure} 

Almost all theoretical models expect small effects in the
reaction $\pdupp$ in the central region, because $\pi^0$ is supposed to 
be produced mainly from the gluons. The gluons transversal 
spin component is considered to be small. The Anselmino  
prediction \cite{anselm98} for 200~GeV at small $\xf$ is 
presented together with the E704 experimental data ({\bf Fig5b}).

By comparing the presented data with the 
$\pi^0$ asymmetry $A_N \approx -40\%$ in the reaction 
$\pimp$ at 40~GeV at the same kinematic region \cite{protv},  
we may conclude that the asymmetry depends on quark flavour. 
Otherwise we have to suppose significant changes in the interaction dynamics
in the energy range between 40 and 70~GeV.

The asymmetry in  the reaction $\pdupp$  is 
cancelled because of various channel interference 
from a polarized and non-polarized proton and a large contribution 
of gluonic channels. In the $\pi^-\pdup$ collisions
the valence $u$-quark from a polarized proton combining with the valence 
$\bar{u}$-quark from $\pi^-$ gives the main contribution to $\pi^0$ 
production, while other channels are suppressed.

\subsection*{$A_N$ in the reaction $\pimp$ at the polarized target 
fragmentation region.}

\begin{wrapfigure}{R}{7.8cm}
\centering
\vspace*{-0.2cm}
\mbox{\epsfig{figure=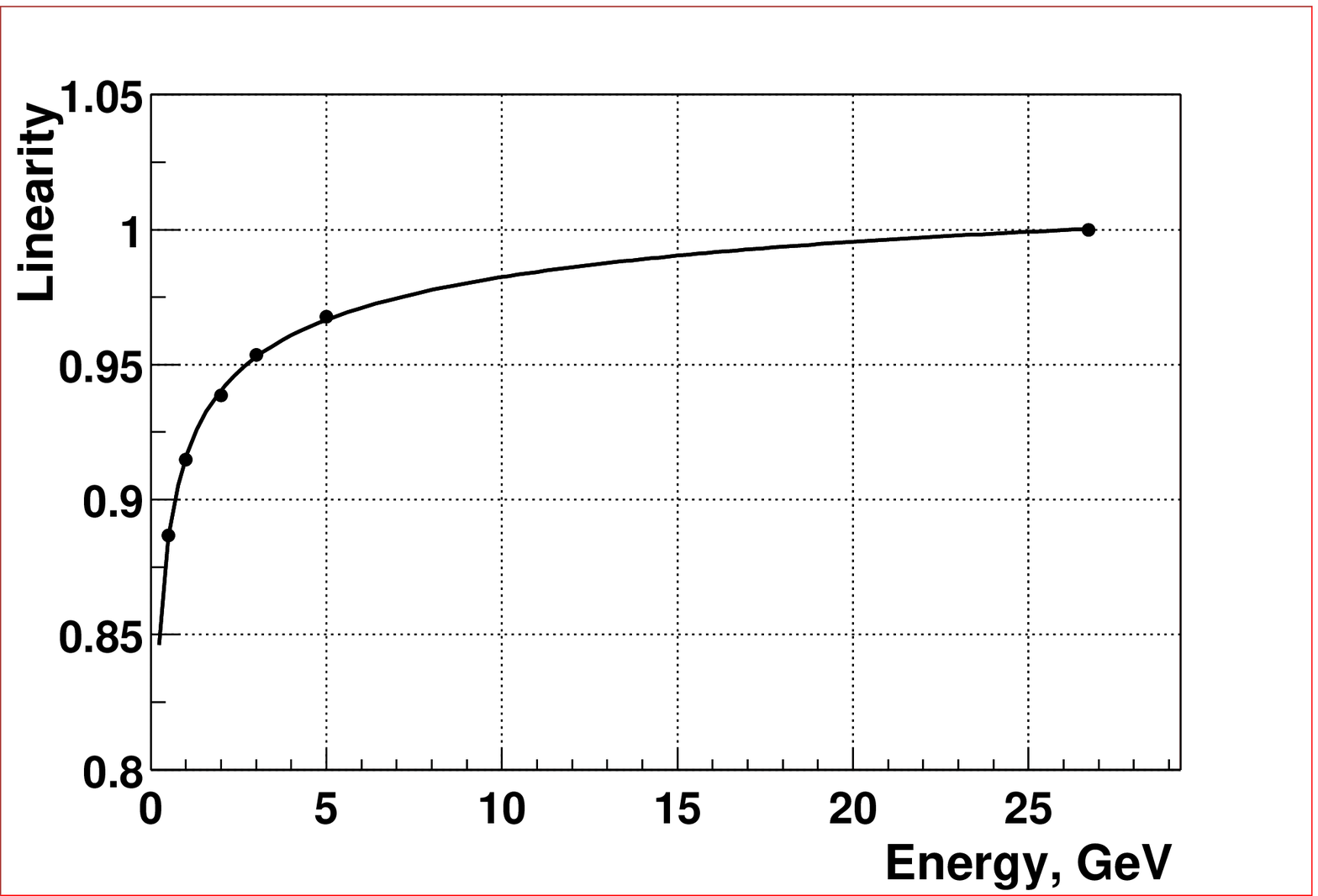,width=7.6cm,
height=4.cm}}\\
\vspace*{0.2cm}
\mbox{\epsfig{figure=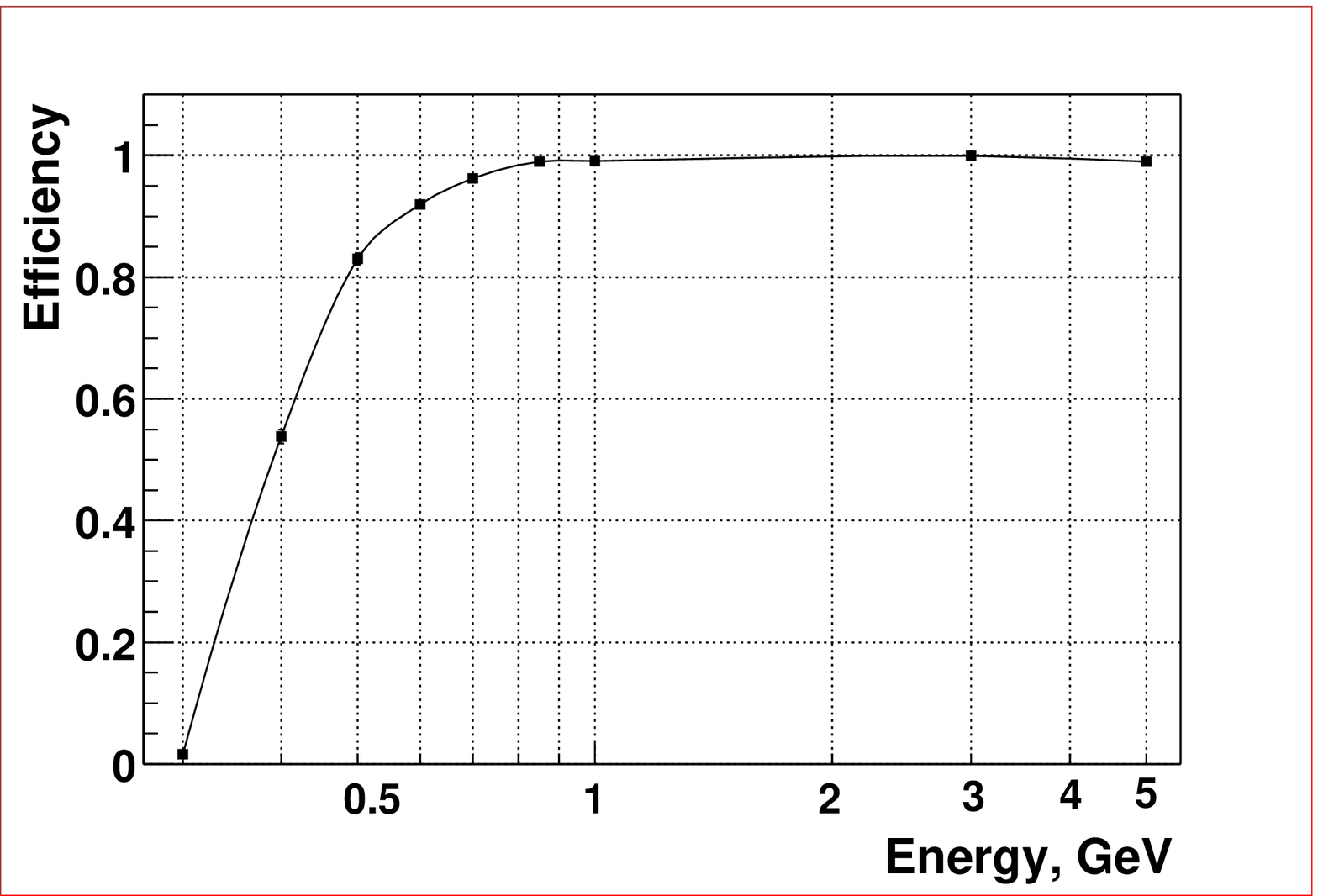,width=7.6cm,height=4.cm}}\\
{\small {\bf Figure 6.} The dependence of the  energy reconstructed (top) and 
the efficiency on true
$\gamma$-quantum energy from MC simulation.
}\\
\vspace*{0.4cm} 
\mbox{\hspace*{1cm} \epsfig{figure=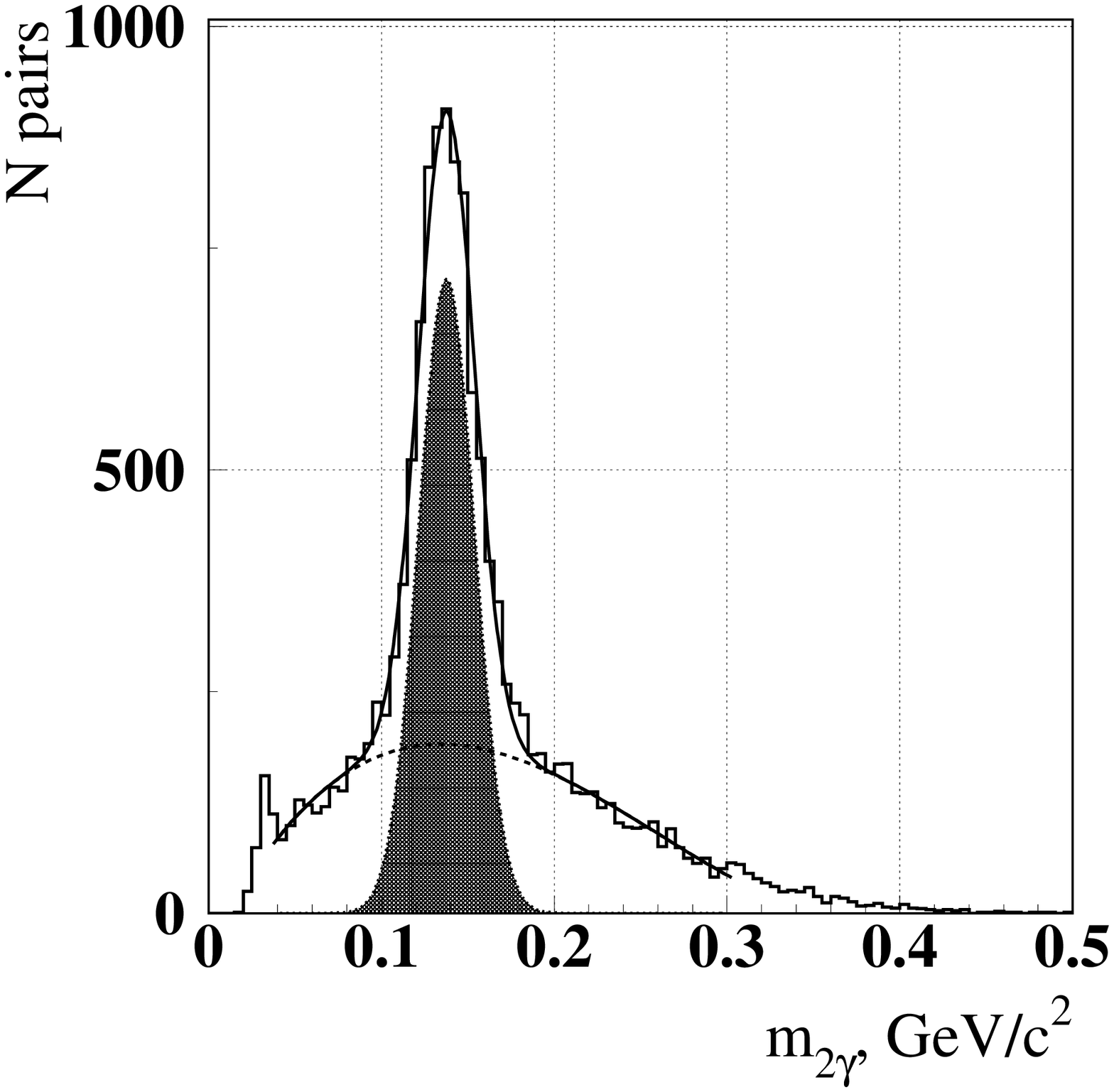,width=5.6cm,
height=5.1cm}}\\
\mbox{\hspace*{1cm} \epsfig{figure=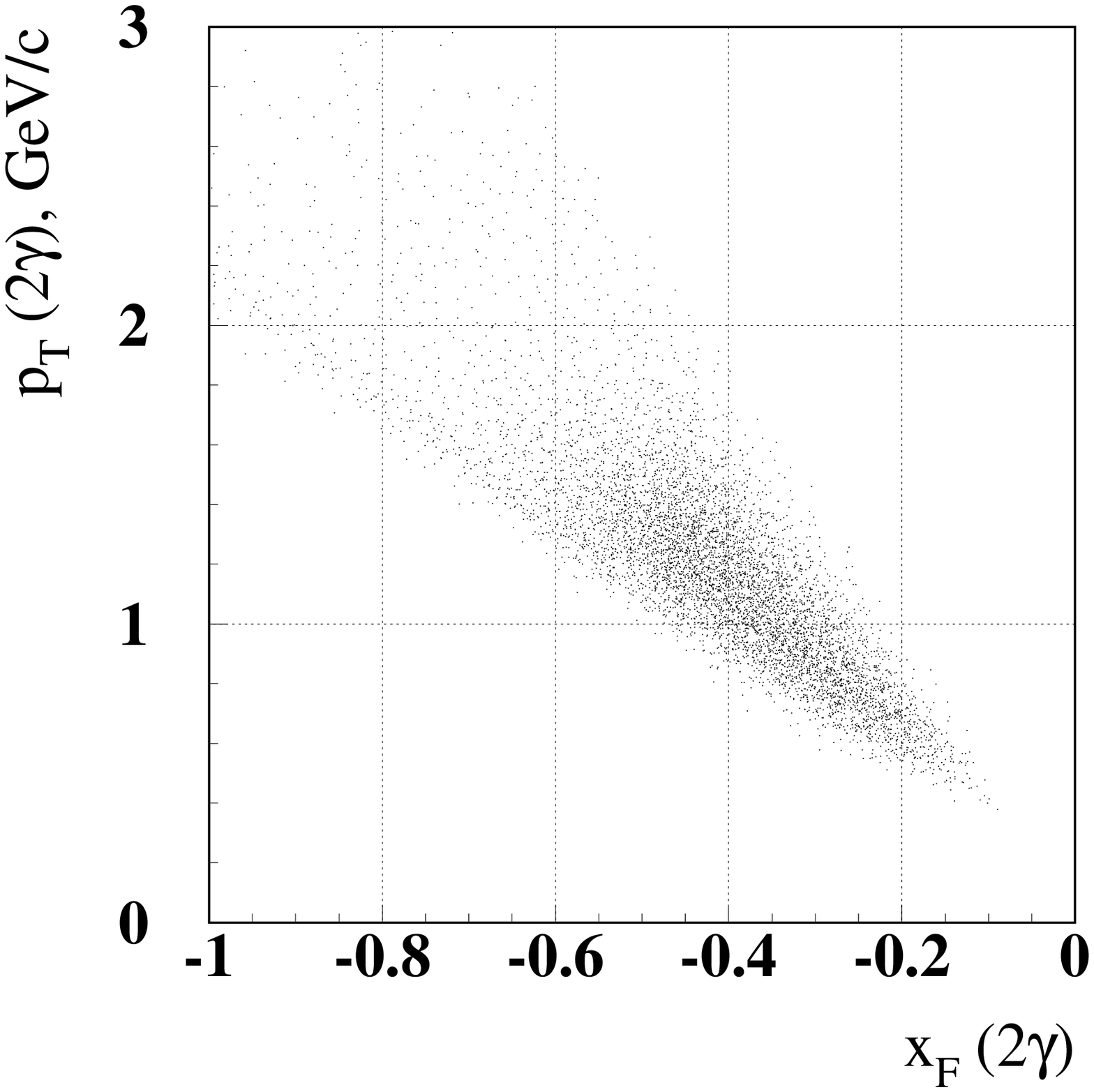,width=5.6cm,
height=5.1cm}}\\
{\small {\bf Figure 7.} $\gamma$ pairs mass spectrum (top) and kinematic
parameters in the $\pi^0$ mass region}
\end{wrapfigure} 

The asymmetry in the polarized target fragmentation 
region was measured for the first time. The aim of the 
current experiment was to check our expectations that the asymmetry
in the polarized particle fragmentation region is the 
same for both the beam and the target projectile. 
The experiment was carried out at the modified experimental 
setup PROZA-M (see above).

The electromagnetic calorimeter from 720 cells was placed at 
2.3~m downstream the target at the angle of $40^{\circ}$ 
in the laboratory frame in two expositions in the 2000 and at the angle
of $30^{\circ}$  in the 1999 data taking run.

We detected $\gamma$-quanta in the energy range between 0.5 and 3.5~GeV.
Monte-Carlo simulation of the electromagnetic showers in the 
PROZA-M lead-glass calorimeter  based on GEANT3.21 
shows that significant 
part of the $\gamma$-quanta energy (up to 20\%) was lost 
during reconstruction procedure mainly due to 
an electronic threshold. Cherenkov light was simulated taking 
into account light absorbtion and reflection from crystal surface 
wrapped by mylar. The dependence of detected energy on 
simulated $\gamma$-quanta energy
is shown in {\bf Fig~6} as well as reconstruction efficiency. 
The efficiency is higher than 80\% starting from the energy 
0.8~GeV. A special algorithm was developed to correct 
these energy losses.

A mass spectrum and the detector kinematic region  
are presented in {\bf Fig.~7}. The $\pi^0$-meson  width 
$\sigma_m = 16$~MeV/$c^2$ and mainly depends on detector 
energy resolution. Kinematic parameters $p_t$ and $\xf$ are correlated.    
It means that the measurement was carried out at rather narrow 
solid angle.

Single armed experimental setup can bring additional
apparature systematic asymmetry displacement. Special
algorithm was developed to eliminate this displacement.
The method was based on the fact that the asymmetry 
of 2$\gamma$-pairs between $\pi^0$ and $\eta$ mass 
is zero \cite{protv}. The method is described in 
the paper \cite{proza40}. 

\begin{wrapfigure}{R}{8cm}
\begin{center}
\vspace*{-1.4cm}
\mbox{\epsfig{figure=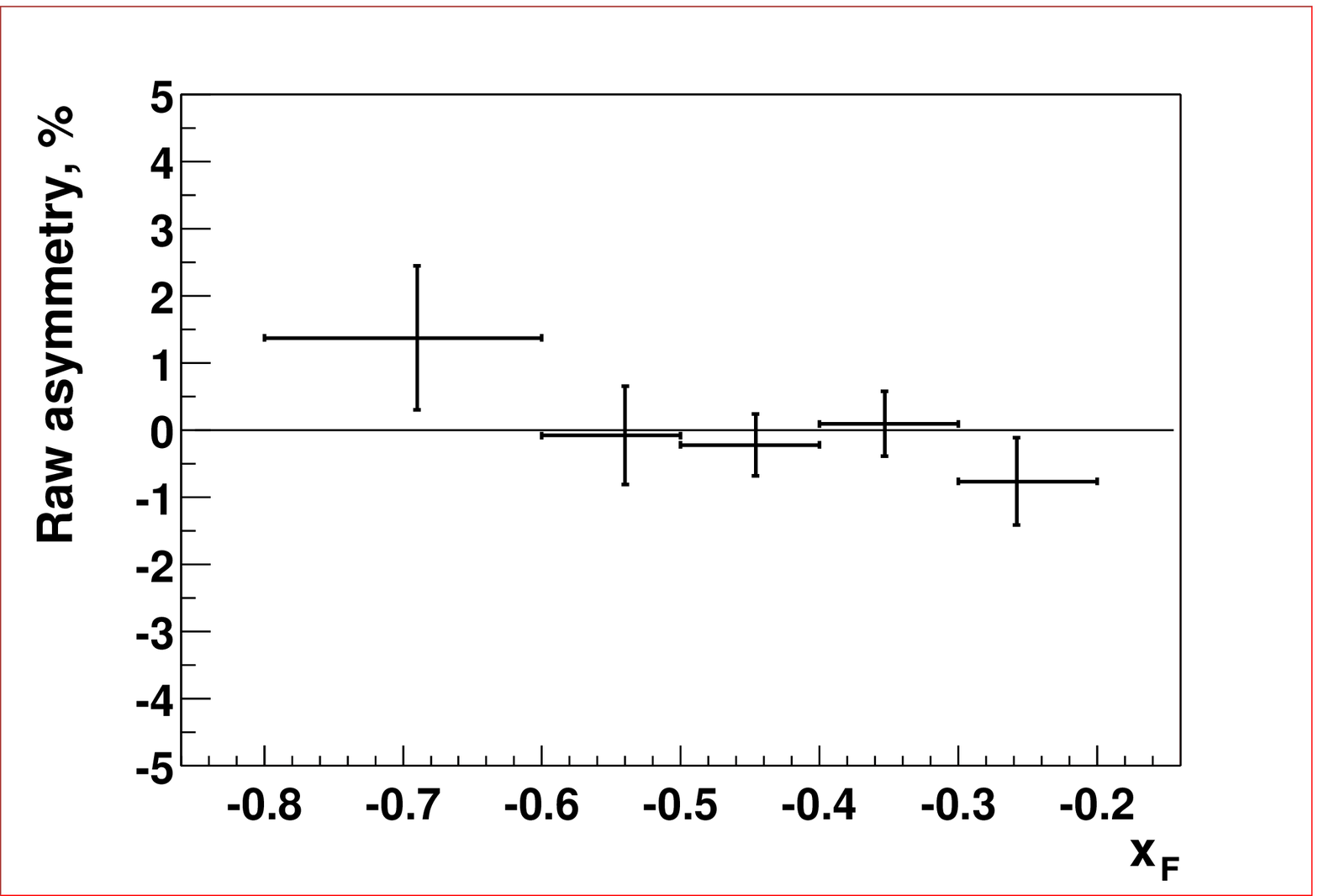,width=7.8cm,height=4.4cm}}\\
\vspace*{0.2cm}
\mbox{\epsfig{figure=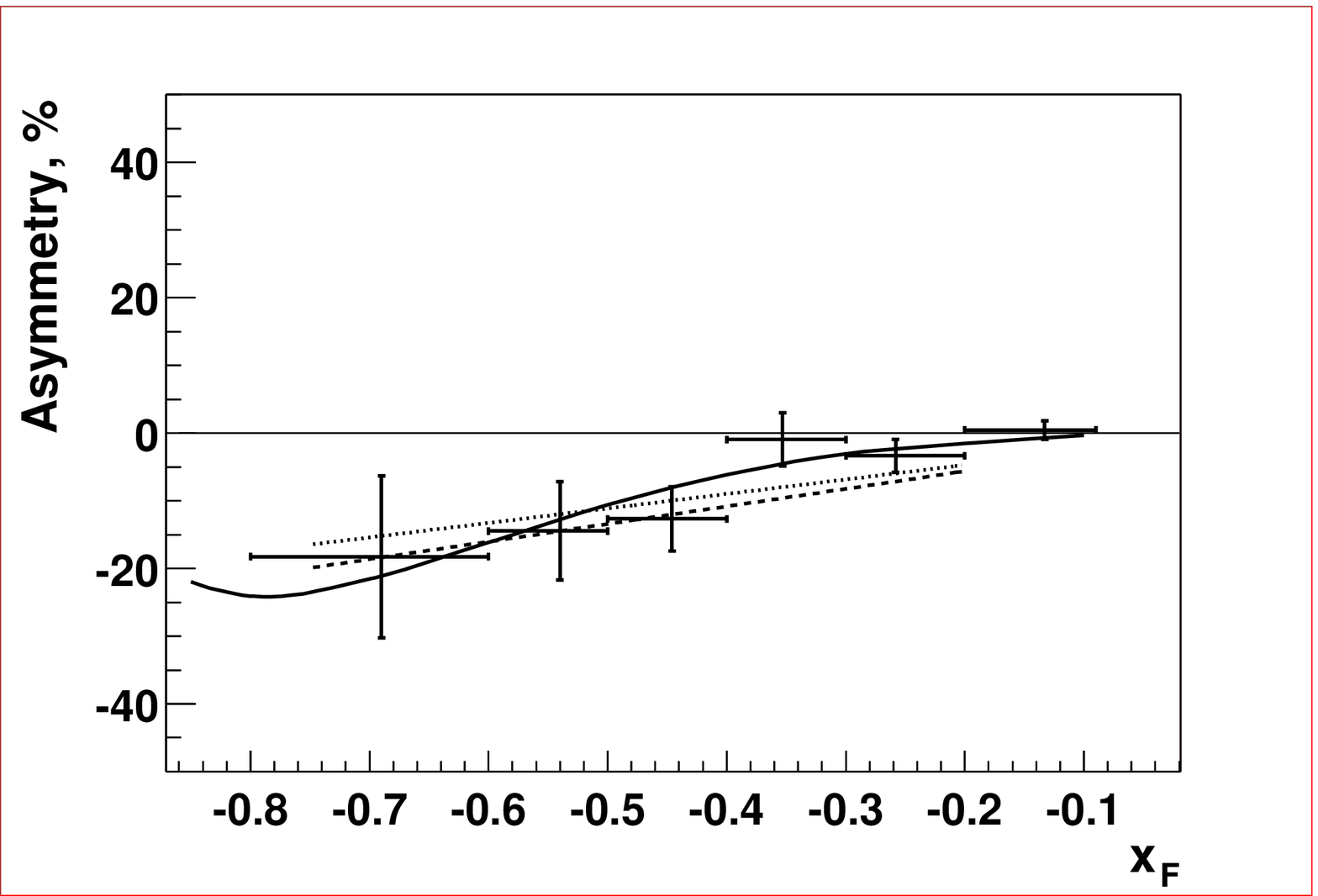,width=7.8cm,height=4.4cm}}\\
\end{center}
{\small {\bf Figure 8.} {\it Top:} Raw false asymmetry.\\
{\it Bottom:} $A_N$ in the reaction $\pimp$ at 40~GeV at the polarized target
frag\-men\-ta\-tion region. Solid line --- Anselmino prediction for Collins
mechanizm; two other lines -- calculation in quark model for U-matrix}
\end{wrapfigure} 

False asymmetry is zero and is presented in {\bf Fig~8 (top)}.

Measured asymmetry $A_N$   
is close to zero at low values of
$|\xf|$ and $p_T$ ({\bf Fig~8 (bottom)}). $A_N=(-13.8 \pm 3.8)\%$ at $-0.8<\xf<-0.4$. 
The result is similar to $\pi^0$ asymmetry in the polarized beam 
fragmentation experiments 
E704 ({($12.4\pm 1.4)\%$}, $\sqrt s=20$~GeV, \cite{e704beam})
and STAR ({$(14\pm 4)\%$, $\sqrt s=200$~GeV \cite{STAR2002}}). 
We can make a conclusion that the asymmetry does not depend
on beam energy at large $\xf$. Analyzing power in this reaction 
can be used for  proton beam polarization measurements.

The result was compared with theoretical predictions.
The Anselmino calculation for the reaction $\pdupp$ is
based on the E704 data and is drawn by solid line in  {\bf Fig~8 (bottom.)}
\cite{ansprivate}. The predictions of the quark model for U-matrix
\cite{troshin} are drawn by dashed and dotted-dashed lines 
\cite{troshprivate} in the same figure.

Earlier the asymmetry in this reaction was measured in
the central region. $A_N$ was found to be small at $p_t<1.6$~Gev/c and 
rose up linearly starting from $p_T \approx 1.65$~GeV/c \cite{protv}.
To compare the results from the two measurements we studied the 
asymmetry dependence on $\pi^0$ energy in the centre of mass system.
Surprisingly the asymmetry  begins to rise up at the same energy 
({\bf Fig~9}).

\subsection*{Conclusions}

Finally  we can summarise:

\begin{wrapfigure}{R}{6cm}
\begin{center}
\vspace*{-2.6cm}
\mbox{\epsfig{figure=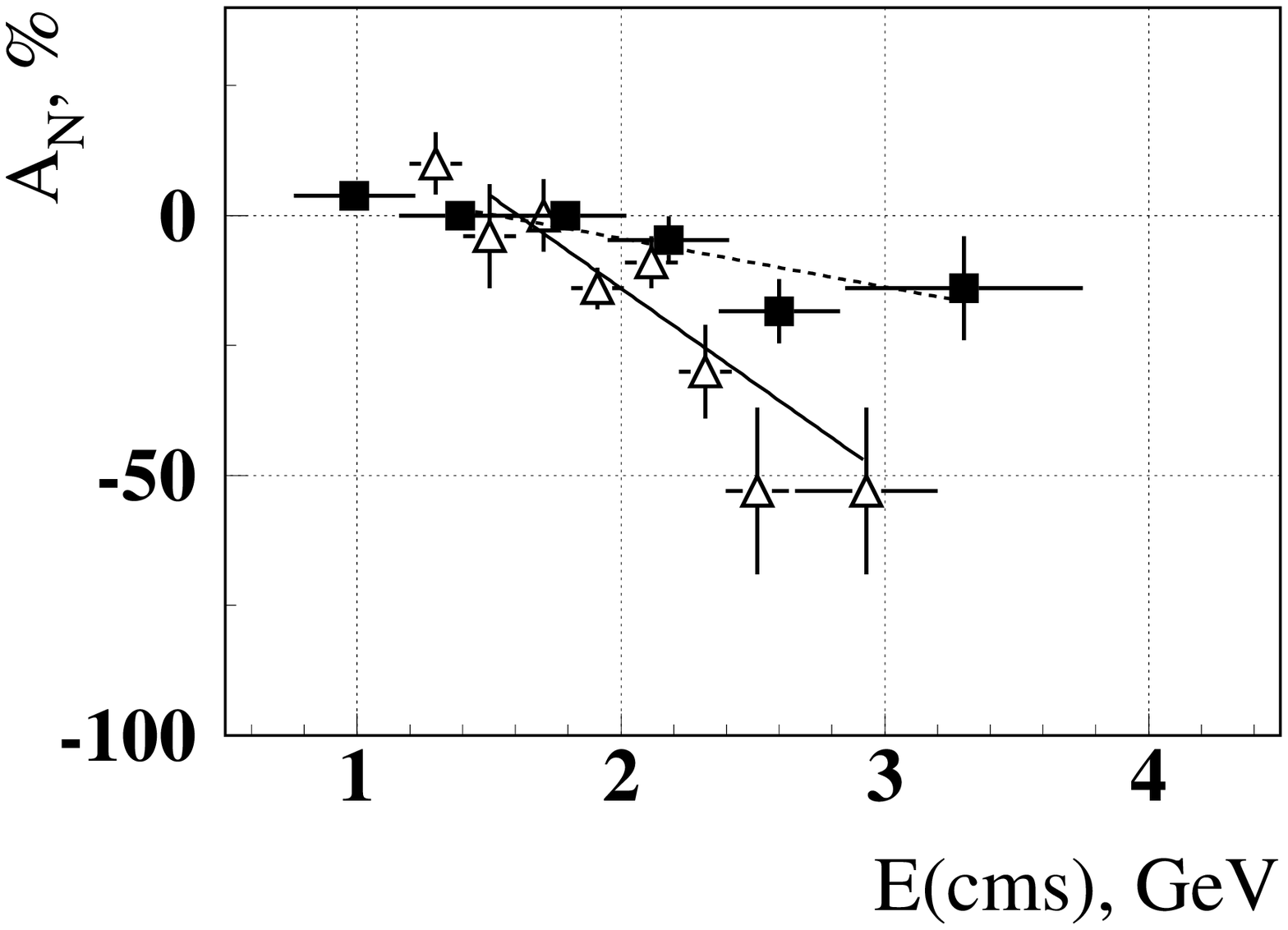,width=5.8cm,height=5.8cm}}\\
\end{center}
\vspace*{-0.4cm}
{\small {\bf Figure 9.} The dependence of $A_N^{\pi^0}$ on the
energy in the centre of mass system in the central region 
($\triangle$, \cite{protv})
and in the target polarization region ($\blacksquare$, current result).}
\end{wrapfigure}

$\bullet$~$A_N$ in  the reaction {$p + p_{\uparrow} \rightarrow \pi^0 + X$} 
is zero within error bars in the central region at $1.0<p_T<3.0$~GeV/c. 
The result is in a good agreement with the E704 measurements at 
200~GeV and differs from the CERN 24~GeV data.

$\bullet$~By comparing presented asymmetry with the measurements 
in the reaction
$\pimp$, we can conclude that the asymmetry depends on quark flavour in the
central region. Another possibility is that the interaction dynamics 
changes dramatically in the energy range between 40 and 70~GeV.

$\bullet$~The asymmetry in the polarized target fragmentation region 
was measured for the first time in the reaction $\pimp$. 
$A_N=(-13.8 \pm 3.8)\%$ at $-0.8<\xf<-0.4$ and $p_T$ 
range between 1 to 2~GeV/c and is close to zero at $-0.4<x_F<-0.1$ 
and $p_T$ from 0.5 to 1.5~GeV/c.

$\bullet$~Asymmetry at large $|\xf|$ is in a good agreement with
E704 (200 GeV) and BNL (equivalent to 20 TeV in the laboratory system) data
in the reaction $\pdupp$ in the polarized beam fragmentation region. 
It was experimentally established, that $A_N$ appears
in the polarized particle fragmentation region (independent on
beam or target). Analysing power of $\pi^0$ inclusive production
can be used for the measurements of the proton beam polarization.

$\bullet$~By comparing the asymmetry in the reaction $\pimp$ at 40~GeV in the 
central region and in the polarized target fragmentation region 
we found that the asymmetry starts to rise up at 
$E_{cms}^0 (\pi^0) \approx 1.7$~GeV/c in the centre of mass system for the 
two different kinematic region. 

{\it The work is supported by RFBR grant 03-02-16919}

\end{document}